\documentclass[aps, prb, reprint, superscriptaddress, amsmath, amssymb, floatfix]{revtex4-1}
\usepackage{xcolor}
\usepackage{graphicx}
\usepackage{dcolumn}
\usepackage{bm}
\usepackage[section]{placeins}

\newcommand{\angstrom}{\mbox{\normalfont\AA}}
\newcommand{\ket}[1]{|#1\rangle}

\newcommand{\avg}[1]{\langle #1 \rangle}

\newcommand{\GW}[0]{$\text{G}_0\text{W}_0\text{ }$}
\newcommand{\dprod}[2]{\langle #1 | #2 \rangle}
\def\mark#1{{#1}}

\begin{document}


\title{Towards fully automatized GW band structure calculations:\\What we can learn from 60.000 self-energy evaluations
}


\author{Asbjørn Rasmussen}
\email{asbra@fysik.dtu.dk}%
\affiliation{CAMD, Department of Physics, Technical University of Denmark, DK-2800 Kongens Lyngby, Denmark}%
\affiliation{Center for Nanostructured Graphene (CNG), Technical University of Denmark, DK-2800 Kongens Lyngby, Denmark}%
\author{Thorsten Deilmann}
\email{thorsten.deilmann@wwu.de}%
\affiliation{%
Institut f\"ur Festk\"orpertheorie, Westf\"alische Wilhelms-Universit\"at M\"unster, 48149 M\"unster, Germany%
}%
\author{Kristian S. Thygesen}
\affiliation{CAMD, Department of Physics, Technical University of Denmark, DK-2800 Kongens Lyngby, Denmark}%
\affiliation{Center for Nanostructured Graphene (CNG), Technical University of Denmark, DK-2800 Kongens Lyngby, Denmark}%


\date{\today}

\begin{abstract}
We analyze a data set comprising 370 GW band structures of two-dimensional (2D) materials covering 14 different crystal structures and 52 chemical elements. The \mark{band structures contain a total of 61716 quasiparticle (QP) energies} obtained from plane wave based one-shot G$_0$W$_0$@PBE calculations with full frequency integration. We investigate the distribution of key quantities, like the QP self-energy corrections and QP weights, and explore their dependence on chemical composition and magnetic state. The linear QP approximation is identified as a significant error source and we propose schemes for controlling and drastically reducing this error at low computational cost. We analyze the reliability of the $1/N$ basis set extrapolation and find that is well-founded with a narrow distribution of coefficients of determination ($r^2$) peaked very close to 1. Finally, we explore the accuracy of the scissors operator approximation and conclude that its validity is very limited. Our work represents a step towards the development of automatized workflows for high-throughput \GW band structure calculations for solids.
\end{abstract}

\maketitle


\section{Introduction}
In computational materials science, the high-throughput mode of operation is becoming increasingly popular\cite{curtarolo2013high}. The development of automatized workflow engines capable of submitting, controlling and receiving thousands of interlinked calculations\cite{jain2015fireworks, pizzi2016aiida, mortensen2020myqueue} with minimal human intervention has greatly expanded the range of materials, and properties, that can be investigated by a single researcher. Several high-throughput studies have been conducted over the past decade mostly with the aim of identifying new prospect materials for various applications including catalysis\cite{greeley2006computational}, batteries\cite{ kirklin2013high,zhang2019computational}, thermoelectrics\cite{ chen2016understanding, bhattacharya2015high}, photocatalysts\cite{castelli2012computational}, transparent conductors\cite{hautier2013identification}, and photovoltaics\cite{ yu2012identification,kuhar2018high}, just to mention some. The vast amounts of data generated by such screening studies have been stored in open databases\cite{ thygesen2016making,saal2013materials,jain2013commentary,curtarolo2012aflow} making them available for further processing, testing and comparison of methods and codes, training of machine learning algorithms etc. With very few exceptions, the high-throughput screening studies and the generation of materials databases, have been based on density functional theory (DFT) at the level of the generalized gradient approximation (GGA).

While DFT is fairly accurate for structural parameters and other properties related to the electronic ground state, it is well known that electronic band structures, in particular the size of band gaps, are not well reproduced by most xc-functionals\cite{godby1986accurate}.
\mark{This holds in particular for the LDA and GGA functionals, which hugely underestimate band gaps, often by about a factor of 2 or more\cite{huser2013quasiparticle,shishkin2007self}. Hybrid functionals and certain metaGGAs perform significantly better\cite{borlido2020exchange}, but are not fully ab-initio and miss fundamental physics such as nonlocal screening effects\cite{garcia2009polarization}.} Instead, the gold standard for quasiparticle band structure calculations of solids is the many-body GW method\cite{hedin1965new,hybertsen1986electron,aryasetiawan1998gw,golze2019gw}, which explicitly accounts for exchange and dynamical screening. In its simplest non-selfconsistent form, i.e. G$_0$W$_0$, this approximation reproduces experimental band gaps to within 0.3 eV (mean absolute error) or 10\% (mean relative error)\cite{shishkin2007self,huser2013quasiparticle,nabok2016accurate}. We note in passing that for partially self-consistent GW$_0$\cite{shishkin2007self} or when vertex corrections are included\cite{shishkin2007accurate,schmidt2017simple}, the deviation from experiments falls below 0.2 eV, which is comparable to the uncertainty of the experimental reference data. The improved accuracy of the GW method(s) comes at the price of a significantly more involved methodology both conceptually and numerically as compared to DFT. While DFT calculations can be routinely performed by non-experts using codes that despite very different numerical implementations produce identical results\cite{lejaeghere2016reproducibility}, GW calculations remain an art for the expert.

The high complexity of GW calculations is due to several factors including: (i) The basic quantities of the theory, i.e. the Greens function ($G$) and screened Coulomb interaction ($W$) are dynamical quantities which depend on time/frequency. Several possibilities for handling the frequency dependence exists including the formally exact direct integration\cite{huser2013quasiparticle} and contour deformation techniques\cite{faber2011first} as well as the controlled approximate analytic continuation methods\cite{caruso2012unified} and the rather uncontrolled but inexpensive plasmon-pole approximations\cite{hybertsen1986electron}. (ii) The formalism involves infinite sums over the unoccupied bands. While most implementations perform the sum explicitly up to a certain cut-off, schemes to avoid the sum over empty states have been developed\cite{umari2010gw,govoni2015large}. (iii) The basic quantities are two-point functions in real space (or reciprocal space) that couple states at different $k$-points. This leads to large memory requirements and makes it unfeasible to fully converge GW calculations with respect to basis set. Consequently, strategies for extrapolation to the infinite basis set limit must be employed\cite{klimevs2014predictive,rasmussen2015computational}. (vi) Unless the GW equations are solved fully self-consistently, which is rarely done and does not improve accuracy\cite{shishkin2006implementation,schmidt2017simple}, there is always a starting point dependence. This has been systematically explored for molecules where it was found that LDA/GGA often comprise a poor starting point whereas hybrids perform better in the sense that they lead to better agreement with experimental ionization potentials and produce more well defined spectral peaks with higher quasi-particle weights\cite{rostgaard2010fully,bruneval2013benchmarking}. These and other factors imply that GW calculations not only become significantly more demanding than DFT in terms of computer resources, but they also involve more parameters making it difficult to assess whether the obtained results are properly converged or perhaps even erroneous. 

Successful application of the high-throughput approach to problems involving excited electronic states, e.g. light absorption/emission, calls for development of automatized and robust algorithms for setting the parameters of many-body calculations such as GW (according to available computational resources and required accuracy level), extrapolating the basis set, and assessing the reliability of the obtained results. The first step towards this goal is to analyse and systematize the data from large-scale GW studies. With a similar goal in mind van Setten et al. compared G$_0$W$_0$@PBE band gaps, obtained with the plasmon-pole approximation, to the experimental band gaps. They analyzed the correlations between different quantities and concluded that that G$_0$W$_0$ (with plasmon-pole approximation) is more accurate than using an empirical correction of the PBE gap, but that, for accurate predictive results for a broad class of materials, an improved starting point or some type of self-consistency is necessary. 

In this work we perform a detailed analysis of an extensive GW data set consisting of G$_0$W$_0$@PBE band structures of 370 two-dimensional semiconductors comprising a total of 61716 QP energies. Our focus is not on the ability of the G$_0$W$_0$ to reproduce experiments, i.e. its accuracy, which is well established by numerous previous studies, but rather on the numerical robustness and reliability of the method and the basis set extrapolation procedure. The calculations employ a plane wave basis set and direct frequency integration; thus the use of projector augmented wave (PAW) potentials represents the only significant numerical approximation. We investigate the distribution of self-energy corrections and quasiparticle weights, $Z$, and explore their dependence on the materials composition and magnetic state. By investigating the full frequency dependent self-energy for selected materials we analyse the error caused by the linear approximation to the QP equation and propose methods to estimate and correct this error. We assess the reliability of a plane wave basis set extrapolation scheme finding it to be very accurate with \mark{coefficient of determination}, $r^2$, values above 0.95 in more than 90\% of the cases when extrapolation is performed from 200 eV. Finally, we assess the accuracy of the scissors operator approach, and conclude that it should only be used when average (maximal) band energy errors of 0.2 eV (2 eV) are acceptable.   

\section{Results and Discussion}
\subsection{The \GW data set}
\label{sec:gwdataset}
\begin{figure*}
\centering
\includegraphics[scale=0.63]{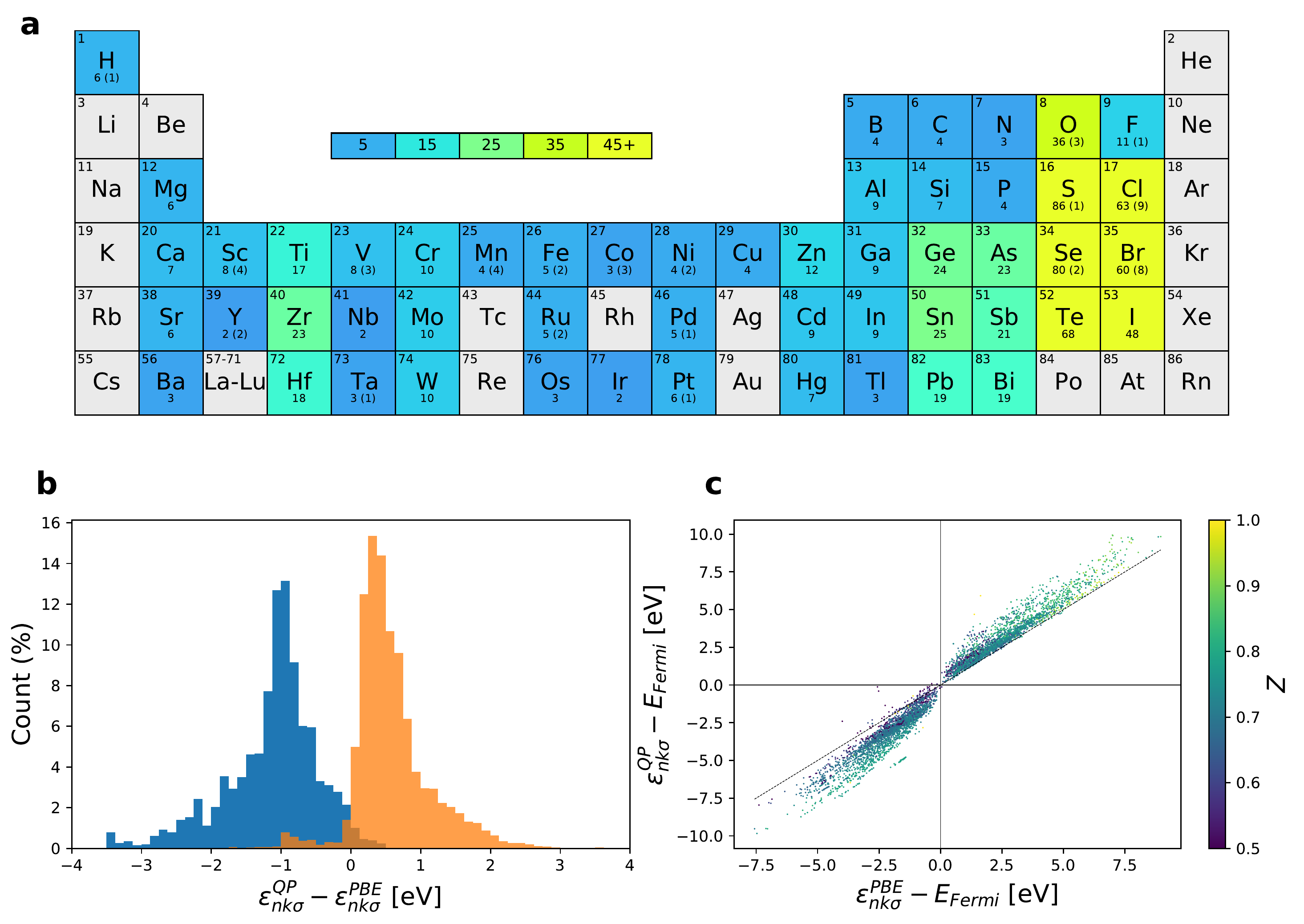}
\caption{(a) The representation of individual elements in the \GW dataset. The number of materials containing a given element is shown under the element's symbol. The number of magnetic materials, if any, is shown in the parenthesis next to the total number. (b) Histograms \mark{of} quasi-particle energy corrections calculated from \GW. The blue histogram show the 3 topmost occupied valence bands, while the orange shows the three lowest unoccupied conduction bands. \mark{(c)} A scatter plot of the PBE energy vs. the \GW energy. The colors show the $Z$ value truncated to the interval [0.5, 1.0]. The points are plotted so that a point with smaller $Z$ are plotted on top of a point with larger $Z$ if the two points overlap.}
\label{fig:IntroFig}
\end{figure*}

The 370 \GW calculations were performed as part of the Computational 2D Materials Database (C2DB) project \cite{c2db}. Below we briefly recapitulate the computational details behind the \GW calculations and refer to Ref. \onlinecite{c2db} for more details. All calculations were performed with the projector augmented wave function code GPAW \cite{enkovaara2010electronic}.

The C2DB database contains about 4000 monolayers comprising both known and hypothetical 2D materials constructed by decorating experimentally known crystal prototypes with a subset of elements from the periodic table \cite{c2db}. Currently, \GW calculations have been performed for 370 materials covering 14 different crystal structures and 52 different chemical elements. Fig. \ref{fig:IntroFig}a illustrates the distribution of elements. The number of materials containing a given element is shown below the element symbol. The number of magnetic materials containing the elements is shown in a parenthesis next to the total number.


To give an overview of some of the data analysed in this work, the distribution of the 61716 \GW corrections for the six bands around the band gap is shown in figure \ref{fig:IntroFig} (b). The distribution for the valence bands is shown in blue and for the conduction in orange. It is usually the case in GW studies that the DFT valence bands are shifted down and the conduction bands are shifted up. A similar behavior is found for the main part of our data, but we also observe a small subset of states for which the correction has the opposite sign. \mark{It is difficult to provide a clear physical explanation for why some occupied states are shifted up and some empty states are shifted down. We stress, however, that the GW corrections are measured relative to the PBE band energies, which is a somewhat arbitrary reference. For example, G$_0$W$_0$@LDA and G$_0$W$_0$@HSE would give different results -- not so much for the resulting QP energies, which are relatively independent of the starting point -- but for the size and sign of the GW corrections, which would now be measured relative to the LDA and HSE energies, respectively.}

Figure \ref{fig:IntroFig} (c) shows a scatter plot of the PBE energies versus the \GW energies. We only show energies from -10 eV to 10 eV for clarity. The color of a point shows the $Z$ value. The latter has been truncated to the region [0.5, 1.0] to show the variation of the main part of the distribution. The main observation we can make from this figure is that there is no obvious correlation between the energies and the $Z$ values. This is also verified by the calculated correlation coefficient, $C$, between $E_\text{PBE}$ and $Z$ ($C$ = 0.27), $E_{\text{G}_0\text{W}_0}$ and $Z$ ($C$ = 0.23) and between the \GW correction, $E_{\text{G}_0\text{W}_0} - E_\text{PBE}$, and $Z$ ($C$ = 0.10). We conclude that there is no significant correlation between the energies and $Z$, meaning that low $Z$ values (which signals a break down of the QP approximation) may occur in any energy range.

\subsection{Quasiparticle weight $Z$}
\label{sec:quasiparticleWeights}

\begin{figure*}
\centering
\includegraphics[scale=1.0]{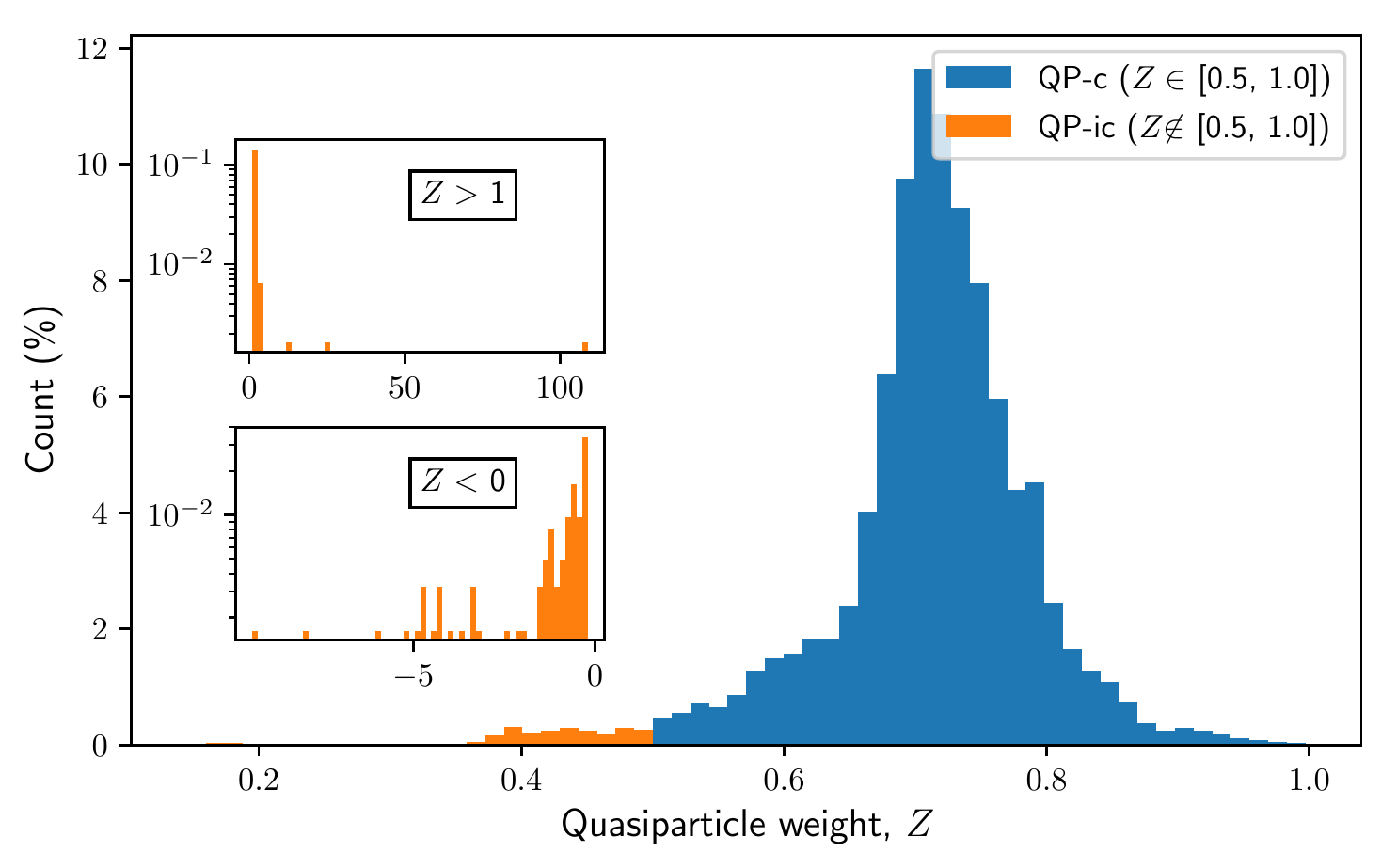}
\caption{Histogram of QP weights, $Z$, for the 61716 QP states in the C2DB \cite{c2db}. The $Z$ values have been extrapolated to the infinite plane wave limit (see next section). The main panel shows the distribution of $Z$ values within the range, $Z \in [0, 1]$, while the upper and lower insets show the distribution outside the physical range, i.e. $Z > 1$ and $Z < 0$, respectively. 0.16\% of points lie in the $Z > 1$ range, while 0.12\% lie in the $Z < 0$ range.}
\label{fig:QuasiparticleWeights}
\end{figure*}

\begin{figure}
    \centering
    \includegraphics[scale=1]{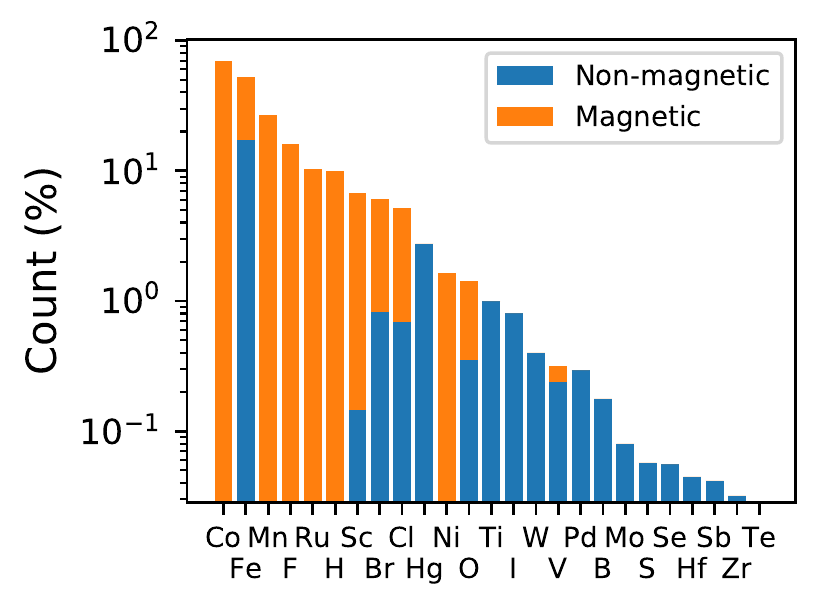}
    \caption{Barplot showing the percentage of QP-ic $Z$ values ($Z \not \in [0.5, 1.0]$) for the given element. Non-magnetic materials are shown in blue and magnetic materials are shown in orange.}
    \label{fig:badZBar}
\end{figure}

\begin{figure*}
    \centering
    \includegraphics[scale=1]{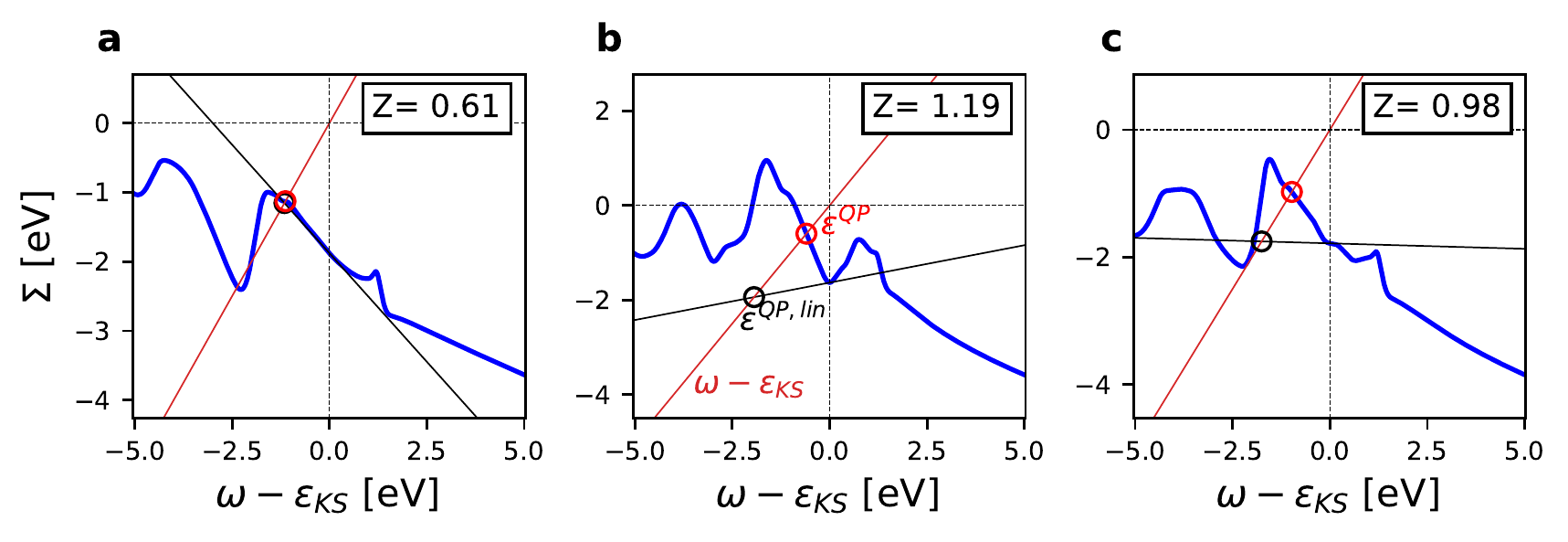}
    \caption{Frequency dependent self-energy (blue) for three electronic states with different quasiparticle weights, $Z$. The red line indicates $\omega - \epsilon_\text{KS}$ while the black line is the linear approximation of the self-energy. The intersection of the blue and red lines indicate the solution to the quasi-particle equation, while the intersection between the red and black lines indicate the solution given by the linear approximation to the self-energy.}
    \label{fig:sigMinimal}
\end{figure*}



The quasiparticle weight, $Z$, gives a rough measure of the validity of the quasiparticle picture, i.e. how well the charged excitations of the interacting electron system can be described by single-particle excitations from the ground state. In the Methods section we prove a physical interpretation of the quasiparticle weight.

In the following we analyse the 61716 calculated QP weights, $Z$, contained in the C2DB database. As discussed in the Methods section, for the QP approximation to be well-founded $Z$ should be close to 1. We split the $Z$ values into two classes: quasiparticle-consistent (QP-c) for $Z \in [0.5, 1.0]$ and quasiparticle-inconsistent (QP-ic) for $Z \not \in [0.5, 1.0]$. \mark{With this definition, QP-c states will have at least half of their spectral weight in the quasiparticle peak, but there is no deeper principle behind the threshold value of 0.5.} We can expect that the QP approximation is more accurate for QP-c states than for QP-ic states.


Figure \ref{fig:QuasiparticleWeights} shows a histogram of the $Z$-values (all  extrapolated to the infinite plane wave limit) corresponding to the 3 highest valence bands and 3 lowest conduction band of 370 semiconductors. 
The vast majority of the values are distributed around $\approx 0.75$ with only 0.28\% lying outside the physical range from 0 to 1 (0.16\% are larger than one and 0.12\% are negative). We find that  97.5\% of the states are QP-c. 

It is of interest to investigate if there are specific types of materials/elements that are particularly challenging to describe by G$_0$W$_0$. Figure \ref{fig:badZBar} shows a barplot of the percentage of QP-ic states in materials containing a given element (note the logarithmic scale). The result of this analysis performed on the non-magnetic (ferromagnetic) materials is shown in blue (orange). For example, a large percentage (about ~65\%) of the states in Co-containing materials are QP-ic. It is clear that magnetic materials contribute a large fraction of the QP-ic states. In fact,  0.36\% of the non-magnetic states are QP-ic while 22\% of the magnetic states are QP-ic. In general, it thus seems that the QP approximation is generally worse for magnetic materials.

\mark{We note that the employed PAW potentials are not strictly norm-conserving. It has previously been found that the use of norm-conserving pseudopotentials can be crucial for the quantitative accuracy of \GW results for materials with localized $d$ or $f$ states. \cite{klimevs2014predictive, jiang2016g, jiang2018revisiting} To investigate this potential issue, we checked the distributions of \GW corrections and QP weights for materials containing at least one element with a pseudo partial wave of norm less than 0.5, i.e. materials where the norm-conservation could potentially be strongly violated for certain states. Out of the 370 materials there were 279 materials in this category. The resulting distributions were not found to deviate qualitatively from those of all the materials (shown in Fig. \ref{fig:IntroFig}(b) and Fig. \ref{fig:QuasiparticleWeights}, respectively), and the strongest indicator of unphysical $Z$ values or opposite-sign \GW corrections remained the magnetic state of the material. On basis of this analysis we conclude that the use of non-norm conserving PAW potentials does not affect the conclusions of our study.}

Based on the distribution of QP weights in figure \ref{fig:QuasiparticleWeights}, it appears that the QP approximation is valid for essentially all the states in the non-magnetic materials and most of the states in the magnetic materials. However, while a QP-c $Z$ value is likely a necessary condition for predicting an accurate QP energy from the linearized QP equation [Eq. (\ref{eq:QPE}) in the Method section], it is not sufficient. This is because the assumption behind Eq. (\ref{eq:QPE}), i.e. that $\Sigma(\varepsilon)$ varies linearly with $\varepsilon$ in the range between the KS energy and the QP energy, is not guaranteed for QP-c states. This is illustrated in figure \ref{fig:sigMinimal} which shows the full frequency dependent self-energy for three states in the ferromagnetic $\text{FeCl}_2$. Case (a) is a typical example where the self-energy of a QP-c state ($Z=0.61$) varies linearly around $\varepsilon_{\mathrm{KS}}$ and the 1st order approximation works well. The second case (b) shows an example where the 1st order approximation breaks down for a QP-ic state ($Z=1.19$). The final case (c) illustrates that the 1st order approximation can break down even in cases where $Z$ is very close to 1.
Unfortunately, there is no simple way to diagnose such cases from the information available in a standard \GW calculation ($\Sigma(\varepsilon_{\mathrm{KS}})$ and $Z$). We stress that the example in figure \ref{fig:sigMinimal}(c) is a special case and that 
in general, the linear approximation is significantly more likely to hold for QP-c states than for QP-ic states (see discussion below).


\subsection{Beyond the linear QP approximation}
\label{sec:beyondlinear}
\begin{figure}
\centering
\includegraphics[scale=1]{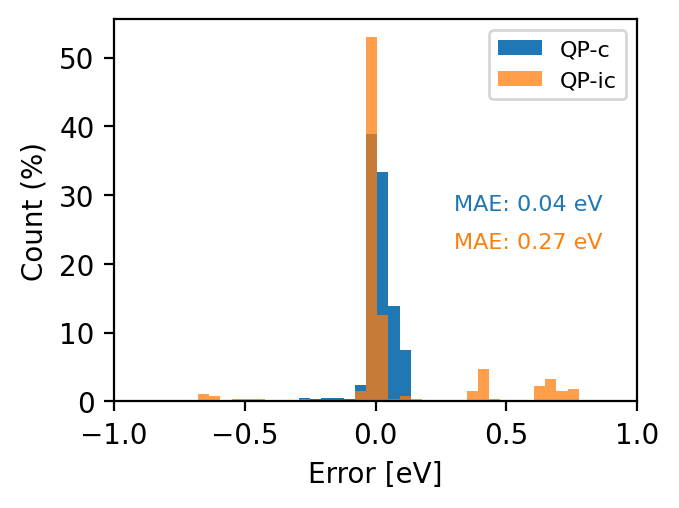}
\caption{The distributions of the error incurred by the linear approximation as estimated from 3192 states in 12 different materials for which we have calculated the full frequency dependent self-energy and determined the exact QP energy (see main text). The distribution for QP-c states is shown in blue, while the distribution for QP-ic states is shown in orange. The inset shows the full distribution for QP-ic states.}
\label{fig:Zbinnederror}
\end{figure}

\begin{figure*}
\centering
\includegraphics[scale=0.7]{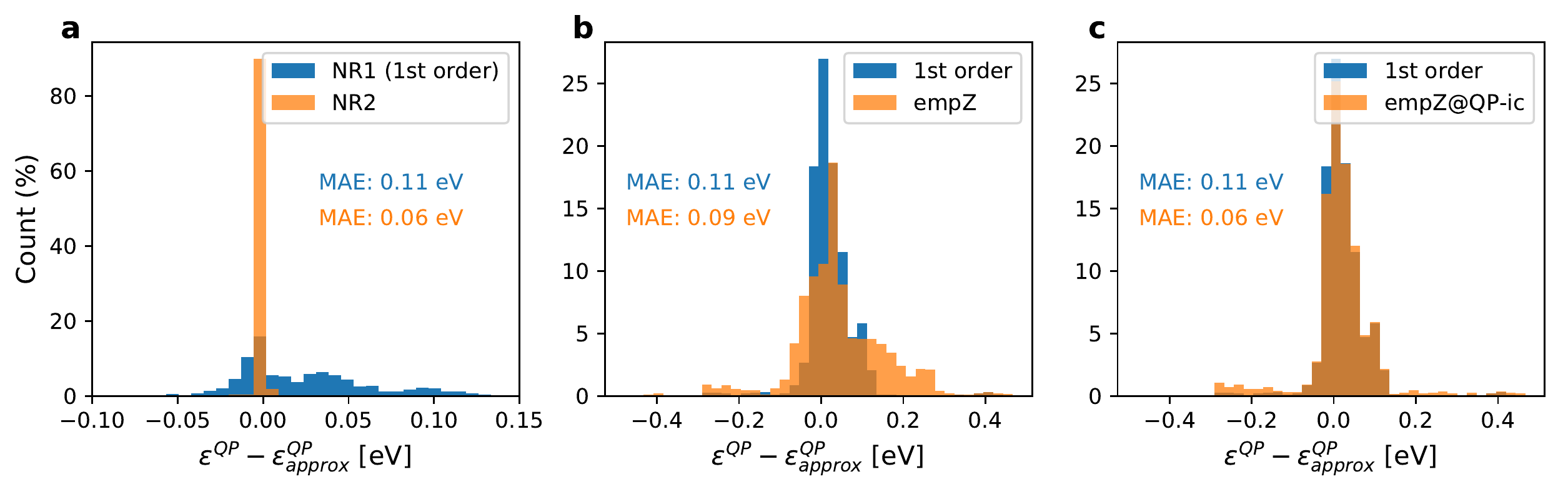}
\caption{(a) The error distributions for first order Newton-Raphson (NR1) (blue) and second order Newton-Rahpson (NR2) (orange). NR1 is equivalent to solving the linearized QP equation. (b) \mark{The NR1 distribution from (a) is again shown in blue for comparison}. The orange distribution shows the error for the empirical empZ scheme. (c) \mark{The NR1 distribution is again shown in blue.} The orange distribution is the error when the empZ scheme is applied only to the QP-ic states.}
\label{fig:emZ}
\end{figure*}

\begin{figure*}
\centering
\includegraphics[scale=0.70]{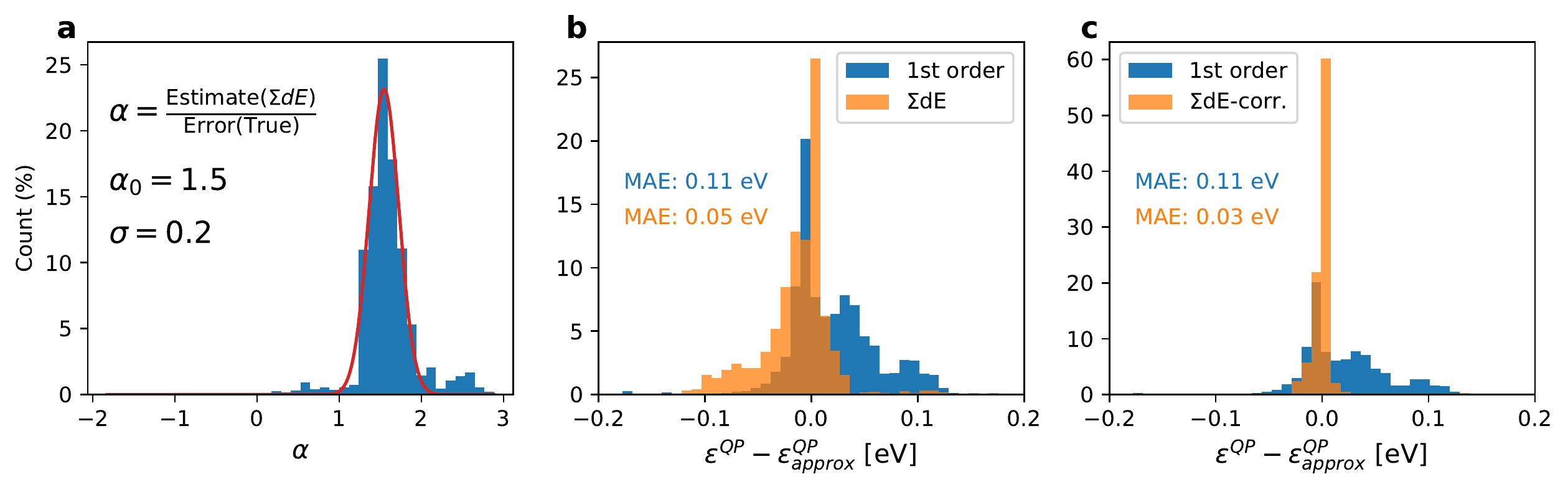}
\caption{(a) The distribution of the ratio of the estimated error and the true error. Also shown in red is a gaussian fit to the distribution. The text annotations shown the definition of $\alpha$ (top), the mean of the fitted gaussian, $\alpha_0$ (middle), and the standard deviation of the fitted gaussian, $\sigma$, (bottom). (b) Distribution of the error of the linear approximation (blue) and the error of solution derived from the estimated error (orange). (c) Correcting for the mean of $\alpha$ yields improved solution estimates (orange).}
\label{fig:sigmadecorrection}
\end{figure*}

\begin{table}[]
\caption{\mark{Summary of the 12 materials used to study the frequency dependent self-energy.}}
\centering
\begin{tabular}{c|c|c|c|c}
    Material & Prototype & Mag. state & PBE gap [eV] & \GW gap [eV] \\
    \hline
    HfBrI & MoSSe & NM & 0.71 & 1.61 \\
    HfClI & MoSSe & NM & 0.81 & 1.78 \\
    ZrBrCl & MoSSe & NM & 0.91 & 1.88 \\
    ZrClI & MoSSe & NM & 0.88 & 1.74 \\
    FeCl$_2$ & MoS$_2$ & FM & 0.35 & 0.00 \\
    MnBr$_2$ & MoS$_2$ & FM & 1.59 & 2.02 \\
    MoS$_2$ & MoS$_2$ & NM & 1.58 & 2.53 \\
    PdSe$_2$ & CdI$_2$ & NM & 0.56 & 1.61 \\
    Al$_2$Se$_2$ & Ga$_2$S$_2$ & NM & 1.99 & 3.54 \\
    Ga$_2$S$_2$ & Ga$_2$S$_2$ & NM & 2.32 & 4.08 \\
    Ga$_2$Se$_2$ & Ga$_2$S$_2$ & NM & 1.76 & 3.44 \\
    In$_2$S$_2$ & Ga$_2$S$_2$ & NM & 1.67 & 3.15
\end{tabular}
\label{fig:materialstable}
\end{table}

\begin{table}[]
\caption{Mean Absolute Errors (MAE) and number of $\Sigma$ evaluations for the various methods discussed in the main text.}
\centering
\begin{tabular}{l|c|c}
    Method & MAE [eV] & \#$\Sigma/Z$ evals \\
    \hline
    1st order & 0.11 & 2 \\
    empZ & 0.09 & 1 \\
    empZ@QP-ic & 0.06 & 2 \\
    $\Sigma$dE & 0.05 & 3 \\
    $\Sigma$dE-corr. & 0.03 & 3 \\
\end{tabular}
\label{fig:summarytable}
\end{table}

Under the assumption that the KS wave functions constitute a good approximation to the QP wave functions, so that off-diagonal elements can be neglected, the solution to the QP equation reduces to solving an equation of the form
\begin{equation}
    \omega-\varepsilon_{\mathrm{KS}}=\Sigma(\omega),
\end{equation}
\mark{where $\Sigma(\omega) = \Sigma_{GW}(\omega) - v_{xc}$ is the frequency dependent self-energy (see Methods).}

In this section we investigate different root-finding schemes to estimate the size of the error introduced by the linear approximation and obtain an improved QP energy.  With high-throughput computations in mind, a good algorithm provides a reasonable balance between computation time (number of $\Sigma$/$Z$ evaluations) and accuracy. To benchmark the different schemes we computed the full frequency dependent self-energy for 3192 states, corresponding to the 3 highest valence bands and 3 lowest conduction bands, for 12 of the 370 2D materials (including two ferromagnetic materials). \mark{The two ferromagnetic materials were chosen at random from materials that had some $Z\not\in [0, 1]$. The remaining 10 materials were chosen at random from materials with all $Z\in [0,1]$ and typical $Z$ distributions. An overview of the materials is shown in table \ref{fig:materialstable}.} The self-energy is evaluated on a uniform frequency grid and interpolated using cubic splines. The ``true'' solution of the QP equation is then determined and used to evaluate the errors of the approximate schemes. \mark{In the cases where there are multiple solutions, the smallest correction is selected.}

\mark{We first determine the errors introduced by the linear approximation. Histograms of the errors for QP-c and QP-ic states are shown in figure \ref{fig:Zbinnederror}. This shows that QP-ic generally have larger error and thus warrant particular attention.}

We first consider the iterative Newton-Raphson (NR) method where we limit ourselves to 1 and 2 iterations to keep the number of self-energy evaluations and thus the computational cost low. We note that 1 iteration (NR1) is equivalent to the linear approximation. The distribution of the errors is shown in figure \ref{fig:emZ} (a). Although 87\% of the errors from NR1 are below 0.1 eV, the mean absolute error (MAE) is 0.11 eV due to outliers. Most of these errors are significantly reduced by performing one more iteration of Newton-Raphson (NR2), but again \mark{outliers increase the MAE}. If we evaluate the MAE without the outliers (those lying outside the displayed error range), the MAE reduces to only 0.006 eV.

Motivated by the relatively narrow distribution of $Z$ values in Figure \ref{fig:QuasiparticleWeights} we consider an empirical solution estimate consisting of replacing the actual $Z$ value with the mean value of the distribution, i.e. we simply set $Z=0.75$. This has the advantage of being simple, computationally cheap, and robust in the sense of avoiding outlier $Z$-values arising from local irregularities in $\Sigma$ at the KS energy, see figure \ref{fig:sigMinimal}(b). The resulting error distribution is shown in \ref{fig:emZ} (b). While the central part of the distribution is slightly broadened compared to the 1st order approximation, the MAE is reduced due to a reduction of outliers (enhanced robustness). As shown in panel (c), the central part of the distribution can be narrowed by applying the empirical approach only for QP-ic states, i.e. when $Z\not \in [0.5,1]$. In fact, this approach (empZ@QP-ic) has a MAE equal to that of NR2 but with half the computational cost (two $\Sigma/Z$ evaluations compared to four).    

Next, we examine polynomial fitting of the self-energy. We construct second and fourth order polynomials, $P_n(\omega)$, from the self-energy at energies in a range of $\pm 1$ eV around the KS energy. The cost of the second and fourth order fits are equivalent to three and five self-energy evaluations, respectively. In general, the polynomial fits have rather low correlation coefficients of $C < 0.9$ and are sensitive to the choice of frequency points and self-energy data used for the fit. As a consequence the resulting errors are large (not shown) and the approach is not suitable. We attribute this to our observation that self-energies are often irregular (on the relevant scale of 1 eV) and not well-described by low-order polynomials.

Finally, we consider a scheme that we refer to as $\Sigma$dE, which estimates the error as 
\begin{align}
\label{eq:sigdE}
    \delta =&\Sigma(\varepsilon^\text{QP, lin}) \nonumber \\
    & - \left(\Sigma(\varepsilon^\text{KS}) + \frac{d\Sigma}{d\omega}\Big|_{\omega = \varepsilon^\text{KS}} (\varepsilon^\text{QP, lin} - \varepsilon^\text{KS})\right).
\end{align}
The motivation for this expression is the following. If the linear approximation is exact, then $\delta$ vanishes as it should. Moreover, if the self-energy has a non-zero curvature it can be shown that $\delta$ equals the true error to leading order in the curvature. In that sense it is similar to the second order polynomial fit, but with the important difference that whereas the polynomial fit was based on uniformly distributed points, $\Sigma$dE uses the value and slope at $E^\text{KS}$ and the value at $E^\text{QP, lin}$. 

\mark{In Figure \ref{fig:sigmadecorrection}(a) the distribution of the ratios of the estimated error and true error is shown and t}he errors resulting from Eq. (\ref{eq:sigdE}) are shown in \ref{fig:sigmadecorrection}(b). Compared to the linear approximation, the $\Sigma$dE reduces the MAE from 0.11 eV to 0.05 eV, at the cost of one additional self-energy evaluation. Interestingly, Eq. (\ref{eq:sigdE}) systematically overestimates the error \mark{as shown in} Figure \ref{fig:sigmadecorrection}(a). A Gaussian fit to the distribution (red curve) has a mean value of $\alpha_0=1.5$ and standard deviation of 0.2. \mark{Since the distribution of $\alpha$ is fairly narrow, it is tempting to correct for the systematic error using $\alpha=\alpha_0$, i.e. replacing $\delta \rightarrow \delta / \alpha_0$. We denote this estimate as $\Sigma$dE-corrected. To verify this procedure we randomly bisect the data into a "training" and a "test" set of equal size. $\alpha_0$ is determined from the training set and the MAE is calculated on the test set. The MAEs thus found were always $0.02-0.03$ eV. We performed the same analysis using different sizes of the training set and found that a MAE of $~0.03$ eV is robust with a training set down to $\geq$5\% of data points. This indicates the approach is insensitive to data used to determine $\alpha_0$. In Figure \ref{fig:sigmadecorrection}(c) the $\Sigma$dE-corrected values are shown, where $\alpha_0$ was determined from the full distribution for simplicity.} The $\Sigma$dE\mark{-corrected} scheme shows excellent performance with an almost four-fold reduction of the MAE from 0.11 eV for the linear approximation to only 0.03 eV at a computational overhead of just one additional self-energy evaluation.

\mark{The performance of the different correction schemes are summarized in Table~\ref{fig:summarytable}.}

\subsection{Plane wave extrapolation}
\label{sec:planewaveCutoffExtrapolation}

\begin{figure*}
\centering
\includegraphics[scale=0.4]{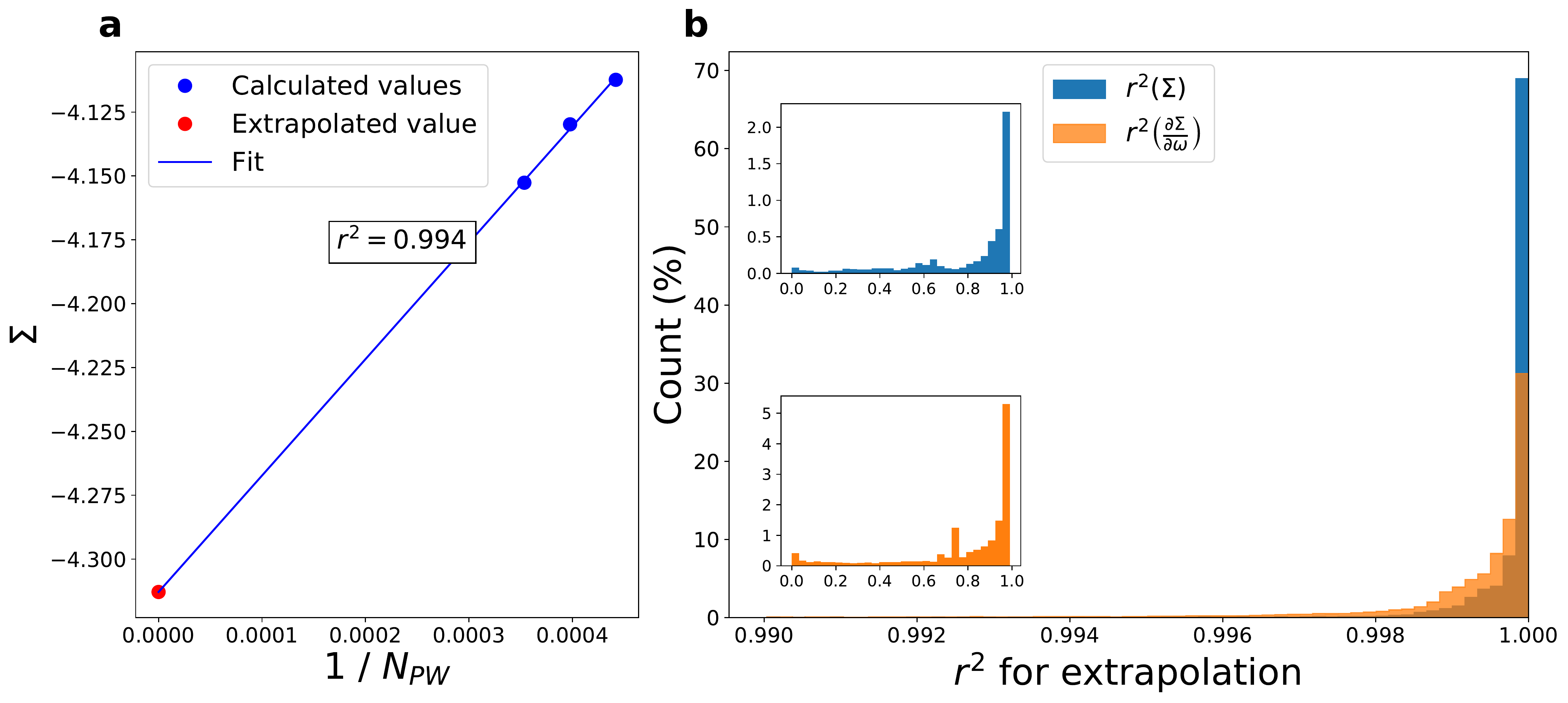}
\caption{(a) Example of the plane wave extrapolation procedure for the \GW self-energy and its derivative. The quantity of interest, e.g. the self-energy, is calculated for three different cutoff energies, here 170 eV, 185 eV, and 200 eV, and the assumed linear dependence on $1/N_{\mathrm{PW}}$ ($N_{\mathrm{PW}}$ is the number of plane waves) is extrapolated to the infinite basis set limit. The coefficient of determination for the fit, $r^2$, is shown in the box. (b) Histogram of coefficient of determination, $r^2$, for the 61716 plane wave extrapolations of self-energies (blue) and the derivatives of the self-energy (orange). The plot shows the distribution for the coefficient of determination $r^2 \geq 0.99$, while the insets show values outside this range. A total of 5.5\% and 14.1\% of the values are $<0.99$ for the self-energy and its derivative, respectively.}\label{fig:sigmaR2Histo}
\end{figure*}

The self-energy and the derivative of the self-energy (both evaluated at the KS energy) are calculated at three cutoff energies: 170 eV, 185 eV, and 200 eV. These values are then extrapolated to infinite cutoff, or infinite number of plane waves, \mark{$N_\mathrm{PW} \rightarrow \infty$}, by assuming a linear dependence on the inverse number of plane waves \cite{rozzi2006exact}. An example of this fitting procedure is shown in figure \ref{fig:sigmaR2Histo} (a).
The extrapolation procedure saves computational time while improving the accuracy of the results - provided the extrapolation is sufficiently accurate. Extrapolation can fail if convergence as a function of plane wave cutoff for the given quantity does not follow the expected $1/N_{\text{PW}}$ behaviour in the considered cutoff range.

To validate this approach we investigate the distribution of the $r^2$ values for all 61716 extrapolations in C2DB. We split them into two cases: extrapolation of the self-energy and extrapolation of the derivative of the self-energy. The distributions are shown as histograms in figure \ref{fig:sigmaR2Histo} (b). The distributions are clearly peaked very close to 1, and in general it seems that the extrapolation is very good. The distribution for the derivatives is somewhat broader, and the extrapolation is generally less accurate than for the self-energies, which indicates a slower convergence with plane waves than for the self-energies. If we choose $r^2 = 0.8$ as an acceptable threshold, we find that 1.7\% of the $r^2$ values of the self-energy extrapolation fall below this criterion while 5.0\% are below for the derivative extrapolation. While these numbers might seem large, the problem is readily diagnosed (by the $r^2$ value) and can be alleviated by using higher plane wave cutoffs.


\subsection{Scissors operator approximation}
\label{sec:scissorOperatorApproximation}

\begin{figure*}
    \centering
    \includegraphics[scale=0.7]{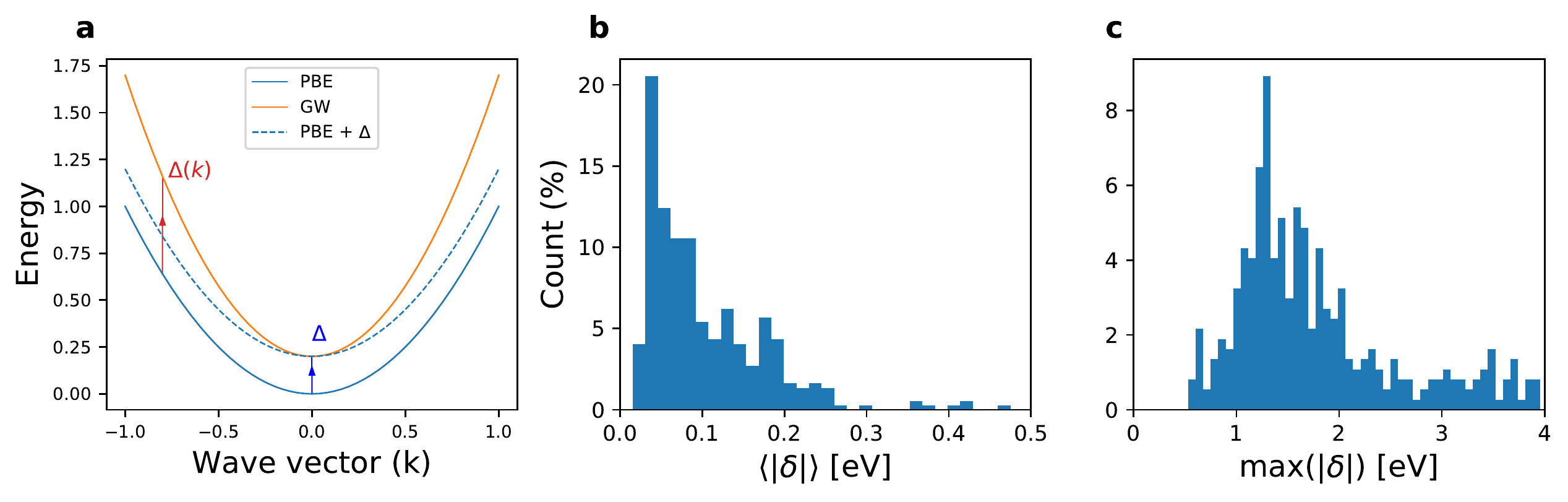}
    \caption{(a) Illustration of the scissors operator approximation for a generic band. The \GW correction ($\Delta$) is calculated at e.g. the $\Gamma$-point and is used to correct the energies at all every k-point. This yields the scissors shifted band structure, here labelled "PBE + $\Delta$". The actual \GW correction at the point $k$ is labelled $\Delta(k)$. (b) Histogram showing the mean absolute error. (c) Maximum absolute error (b) of the scissors operator approximation. In both (b) and (c) the average (maximum) is taken over the 3 highest valence bands and 3 lowest conduction bands in each of the 370 2D materials considered in this work. }
    \label{fig:ScissorHistograms}
\end{figure*}



Within the so-called scissors operator approximation (SOA) it is assumed that the \GW correction is independent of band- and $k$-index. Consequently, the \GW correction calculated at e.g. the $\Gamma$ point is applied to all the eigenvalues thus saving computational time as only one \GW correction is required. In figure \ref{fig:ScissorHistograms} (a) the idea is illustrated for a generic band. With the notation from the figure, the SOA consists of setting $\Delta(k)=\Delta$ (or $\Delta_{n\sigma}(k)=\Delta_{n\sigma}$ when more than one band and spin is involved). 

To test the accuracy of the SOA, we evaluate the mean absolute error ($\avg{|\epsilon|}$) and maximum absolute error ($\mathrm{max}(|\epsilon|)$) of the band energies obtained with the SOA for each of the 370 materials:
\begin{align}
    \avg{|\delta|} = \frac{1}{N_\sigma N_k N_n}\sum_{n, k, \sigma} |\Delta_{n\sigma}(k) - \Delta_{n\sigma}|
\end{align}
and 
\begin{align}
    \mathrm{max}(|\delta|) = \mathrm{max}_{n,k,\sigma}\{|\Delta_{n\sigma}(k) - \Delta_{n\sigma}|\}.
\end{align}
The distribution of these errors are shown in figure \ref{fig:ScissorHistograms} (b) and (c). From panel (b) we see that the mean error exceeds 100 meV for about half of all materials - a rather large error, comparable to the target accuracy of the \GW method itself. Furthermore, it follows from (c) that the maximum absolute error is often $0.5 - 1.0$ eV. We conclude that while the average error of the SOA might be acceptable, it can produce significant errors for specific bands and should be used with care. 

\section{Summary and Conclusions}

As high-throughput computations are gaining popularity in the electronic structure community it becomes important to establish protocols for performing various types of calculations in an automated, robust and error-controlled manner. In this work we have taken the first steps towards the development of automated workflows for \GW band structure calculations of solids. With \GW representing the state-of-the-art for predicting QP energies in condensed matter systems, such workflows are essential for continued progress in the field of computational materials design.  

Based on our detailed analysis of 61716 \GW self-energy evaluations for the eigenstates of 370 two-dimensional semiconductors we were able to draw several conclusions relevant to large-scale GW studies. First of all, we found it useful to divide the states into two categories, namely quasiparticle consistent (QP-c) and quasiparticle inconsistent (QP-ic) states defined by $Z \in [0.5, 1.0]$ and $Z \not \in [0.5, 1.0]$, respectively. Importantly, we found that the QP energies obtained from the standard linearized QP equation are significantly more accurate for QP-c states than for QP-ic state. Moreover, we found the fraction of QP-ic states to be much larger in magnetic materials (22\%) than in non-magnetic materials (0.36\%). Thus extra care must be taken when performing \GW calculations for magnetic materials; in particular, such materials might require a special treatment in high-throughput workflows.  

The mean absolute error (MAE) on the QP energies resulting from the linearized QP equation was found to be 0.11 eV. The MAE evaluated separately for QP-c and QP-ic states was 0.04 eV and 0.27 eV, respectively. In comparison, the accuracy of the GW approximation itself (compared to experiments) is on the order of 0.2 eV. It is therefore of interest to reduce or at least estimate the numerical error bar on the QP energies obtained from \GW calculations. We found that an empirical scheme, where we set $Z=0.75$ (corresponding to the mean of the $Z$-distribution) for QP-ic states, reduces the MAE from 0.11 eV to 0.06 eV with no computational overhead. Similarly, the method dubbed the corrected $\Sigma$dE scheme reduces the MAE to 0.03 eV, at the cost of one additional self-energy evaluation. From these studies it seems natural to accompany the QP energies obtained from \GW with estimated error bars derived from one of the these correction schemes. In fact, we have used the empZ@QP-ic method to correct all the GW the band structures in the C2DB database.

Our analysis of the well known and widely used scissors operator approximation shows that the errors introduced on the individual QP energies when averaged over all bands (specifically the 3 highest valence and 3 lowest conduction bands) typically is on the order of 0.1 eV while the maximum error typically exceeds 1 eV. We stress that our scissors operator fits each of the six bands separately using the \GW corrections at the $\Gamma$-point. Thus the errors introduced by the more standard scissors approximation that fits only the band gap, are expected to be even larger. We conclude that the scissors operator should be used with care and only in cases where errors on specific band energies of 1-3 eV are acceptable. 

Finally, \mark{plane-wave} extrapolation scheme was found to be highly reliable for our PAW calculations when applied to cutoff energies in the range 180-200 eV. In fact only 1.7\% (5.0\%) of the self-energy (derivative of self-energy) extrapolations had an $r^2$ below 0.8. However, for the purpose of high-throughput studies it may be prudent to store and make available information on the $r^2$ for the extrapolation so that the quality of the extrapolation can always be examined and improved calculations with higher cutoff can be performed if deemed necessary.

\section{Methods}
\subsection{\GW Calculations}
For the materials considered here, DFT calculation using PBE \cite{perdew1996generalized} were performed using an 800 eV plane-wave cutoff. Spin-orbit coupling is included by diagonalizing the spin-orbit Hamiltonian in the $k$-subspace of the Bloch states found from PBE.

Those materials that have a finite gap and up to 5 atoms in the unit cell are selected for \GW calculations. The QP energies \mark{in C2DB} are calculated for the 8 highest occupied and the 4 lowest unoccupied bands, \mark{however, in this study we only use the 6 bands closest to the Fermi level (3 valence and 3 conduction bands). Furthermore, we only include materials with a PBE gap greater than 0.2 eV as the accuracy of \GW for materials with very small PBE gaps is questionable.} Three energy cutoffs are used: 170 eV, 185 eV, and 200 eV. The results are then extrapolated to infinite energy, i.e. to an infinite number of plane-waves. This extrapolation is done by expressing the self-energies in terms of the inverse number of plane-waves, $1 / N_\text{PW}$, performing a linear fit, and determining the value of the fit at $1 / N_\text{PW} = 0$, see Refs. \onlinecite{invPW1, invPW2}. 

The screened Coulomb interaction entering in the self-energy is calculated using full frequency integration in real frequency space. To avoid effects from the (artificially) repeated layers. A Wigner-Seitz truncation scheme is used for the exchange part of the self-energy \cite{sundararaman2013regularization} and a 2D truncation of the Coulomb interaction is used for the correlation part \cite{rozzi2006exact, ismail2006truncation}. A truncated Coulomb interaction leads to significantly slower $k$-point convergence because the dielectric function strongly depends on $q$ around $q=0$; this is remedied by handling the integral around $q=0$ analytically, see \cite{huser2013, rasmussen2016}. A $k$-point density of $5.0/\angstrom^{-1}$ was used.

The statistical analyses performed here use the data from all spins, $k$-points, and the three highest occupied bands and the three lowest unoccupied bands. In section \ref{sec:quasiparticleWeights} we consider several examples of the full frequency-dependent self-energies for a randomly selected spin, $k$-point, and band combination, subject to some requirements on the quasi-particle weight, $Z$, which are described below.

\subsection{Quasiparticle theory}
\label{sec:qptheory}

The \GW quasi-particle energies are found by solving the quasi-particle equation (QPE) \cite{shishkin2006implementation}:
\begin{align}\label{eq:QPE_2}
E_{nk\sigma}^\text{QP} = \mathrm{Re}\dprod{\psi_{nk\sigma}}{H_\text{KS} + \Sigma(E_{nk\sigma}^\text{QP})|\psi_{nk\sigma}}
\end{align}
Here $\psi_{nk\sigma}$ is the Kohn-Sham wavefunction for band $n$,  crystal momentum $k$, and spin $\sigma$, $H_\text{KS}$ is the single-particle Kohn-Sham Hamiltonian, \mark{$\Sigma(\omega) = \Sigma_\text{GW}(\omega) - v_{xc}$ is the self-energy}, and $v_{xc}$ is the exchange-correlation potential.

Typically, and in C2DB, the QPE is solved via one iteration of the Newton-Raphson method starting from the KS energy, $\epsilon_{nk\sigma}$, which is equivalent to making a linear approximation of the self-energy. This yields the solution
\begin{align}\label{eq:QPE}
  &E_{nk\sigma}^\text{QP} \approx  \epsilon_{nk\sigma} + Z\mathrm{Re}\left[\dprod{\psi_{nk\sigma}}{\Sigma (\epsilon_{nk\sigma})| \psi_{nk\sigma}} \right], \\
  &Z = \left(1 - \frac{\partial \Sigma}{\partial\omega}\middle|_{\omega=\epsilon_{nk\sigma}}\right)^{-1}.
\end{align}
$Z$ is known as the \textit{quasi-particle weight}. The \GW \textit{correction} is defined as the difference between the \GW energy and KS energy, $\Delta E_{nk\sigma} = E_{nk\sigma}^\text{QP}-\epsilon_{nk\sigma}$.

\subsubsection{Physical Interpretation of $Z$}
Following Ref. \onlinecite{huser2013} we provide here a physical interpretation of $Z$. We denote the many-body eigenstates for the $N$ particle system by $\ket{\Psi^N_i}$, where $i$ is the excitation index. An interesting question is how well the state $\ket{\Psi^{N+1}_i}$ can be described as the addition of a single electron to the ground state $\ket{\Psi^N_0}$. In other words, can we find an state $\phi$ such that $\ket{\Psi^{N+1}_i} \approx c^\dagger_\phi \ket{\Psi^N_0}$? The optimal $\phi$ is determined from maximizing the overlap, i.e.
\begin{align}
    \phi = \underset{\varphi}{\mathrm{argmax}}\left(|\dprod{\Psi^{N+1}_i}{c^\dagger_\varphi | \Psi^N_0}|,\, ||\varphi|| = 1\right)
\end{align}
If the maximal overlap is close to 1 the excited many-body state is well approximated by a single-particle excitation. 

It turns out that the square of this maximal overlap is exactly equal to the QP weight $Z$ defined by Eq. (\ref{eq:QPE}) if it is evaluated at the true QP energy and with the true QP wave function rather than at the KS energy and with the KS wave function. Furthermore $Z$ can be shown to be equal to the squared norm of the QP wave function, which is defined as
\begin{align}
    \psi^\text{QP}_i(\mathbf{r}) = \dprod{\Psi^{N+1}_i}{\hat{\psi}^\dagger(\mathbf{r}) | \Psi^N_0}.
\end{align}
For a proof of these results we refer to Ref. \onlinecite{huser2013}. In standard \GW calculations, the self-energy is evaluated at the KS energy using KS eigenstates. In this case, $Z$ is no longer equal to the exact QP weight but only approximates it. If $Z$ deviates significantly from 1, we can only conclude that either 1) the system is strongly correlated so that the QP approximation fails, or 2) the Kohn-Sham energy and/or wave function are a bad approximation to the true QP energy and/or wave function. In either case we would expect that the \GW calculation is problematic and requires special attention.

\section{Data Availability}
The data is available as an ASE\cite{Hjorth_Larsen_2017} database at \url{https://cmr.fysik.dtu.dk/htgw/htgw.html}.

\section{Acknowledgments}
We acknowledge funding from the European Research Council (ERC)  under  the  European  Union’s  Horizon  2020  research and innovation programme (Grant No. 773122, LIMA). The Center for Nanostructured Graphene is sponsored by the Danish National Research Foundation, Project DNRF103. 
TD acknowledges financial support from German Research Foundation (DFG project no. DE 2749/2-1).

\section{Author Contributions}
AR performed the statistical analyses and full, frequency-dependent self-energy calculations. TD performed the \GW calculations. All authors interpreted the analyses and wrote the article.

\section{Competing Interests}
The authors declare no competing interests.


\bibliography{references}

\end{document}